\documentclass[aps,prl,reprint,groupedaddress]{revtex4-2}

\usepackage{graphicx}
\usepackage{amsmath}

\usepackage[svgnames]{xcolor}
\usepackage{hyperref}
\usepackage[inline]{enumitem}

\hypersetup{
 colorlinks,
 linkcolor={red},
 citecolor={blue},
 urlcolor={blue}
}

\usepackage{tikz}
\usepackage{framed}
\definecolor{shadecolor}{gray}{0.9}
\definecolor{shadecolor}{named}{Lavender}
\setlist[itemize]{noitemsep, nosep, itemsep=0pt, topsep=0pt, itemindent = -0.5em}

\begin{document}

\title{
Artificial Intelligence for VAMOS++: From Data Analysis to Intelligent Instrumentation
}
\date{\today}
\author{M. Rejmund}
\author{A. Lemasson}
\author{D. Ramos}
\affiliation{GANIL, CEA/DRF-CNRS/IN2P3, Bd. Henri Becquerel, BP 55027, F-14076 Caen, France} 

\begin{abstract}
%Abstract
\end{abstract}
\maketitle

\subsection{Introduction}

Modern nuclear physics relies on increasingly sophisticated experimental instruments capable of 
measuring rare reaction products with high precision. Among these instruments, the VAMOS++ 
magnetic spectrometer~\cite{Rejmund2011} at GANIL has become a cornerstone for a wide range of 
studies, including nuclear structure investigations, transfer reactions, fission studies, and measurements 
relevant to nuclear astrophysics.

The scientific reach of VAMOS++ stems from its large acceptance and its capability to identify reaction 
products over a broad range of atomic masses, atomic numbers, and atomic charge states. 
These capabilities allow researchers to collect extensive and information-rich datasets. However, 
extracting the full physical information from these measurements is a challenging task. The analysis 
involves reconstructing ion trajectories through a complex magnetic system and combining information 
from multiple detector subsystems. Traditionally, this process requires lengthy iterative procedures, 
extensive expert intervention, and substantial manual tuning.

\begin{figure}
\centering
\begin{shaded*}
\includegraphics[width=1.\columnwidth]{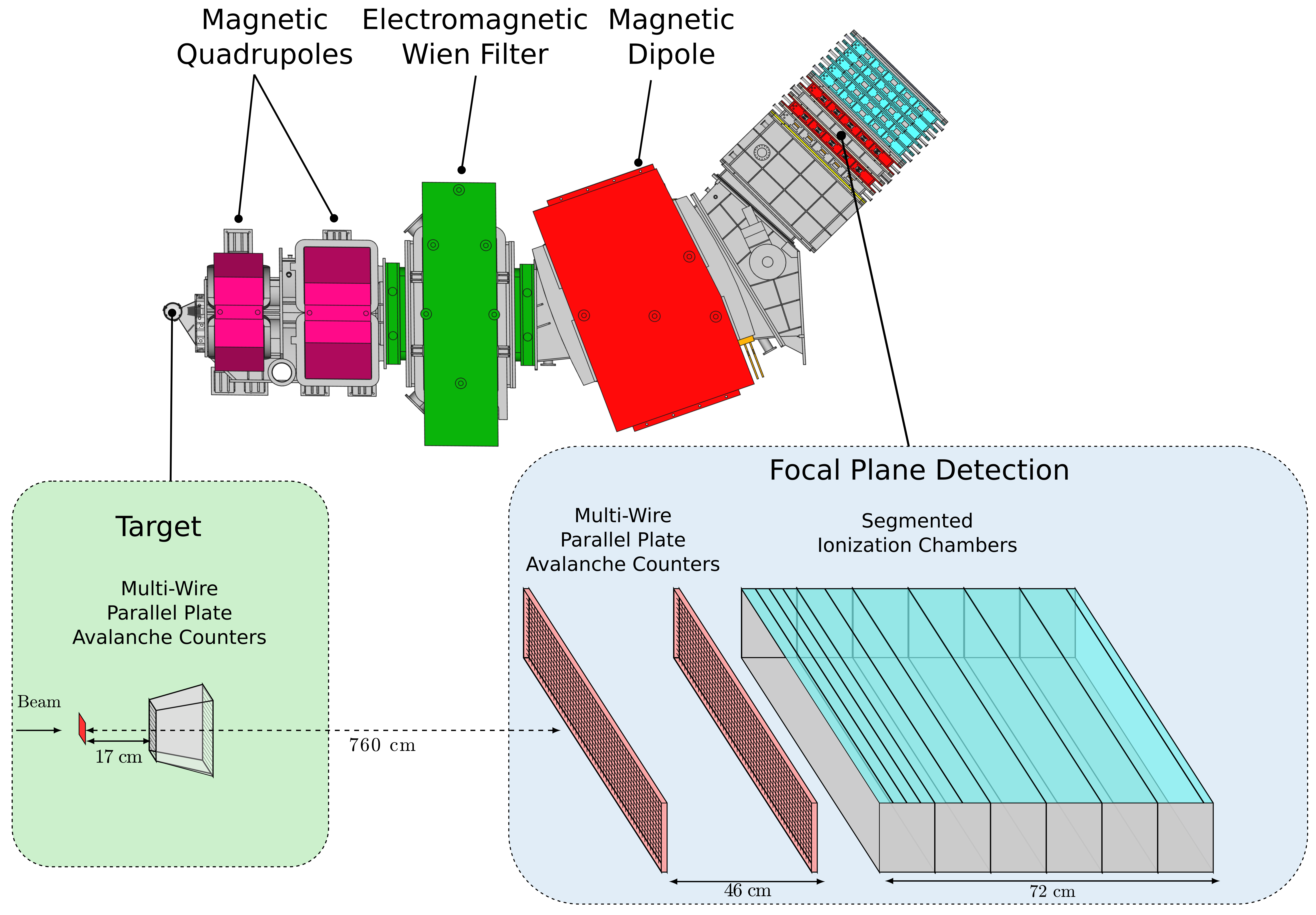}
\caption{
{\bf The VAMOS++ magnetic spectrometer and detection system.}
The large acceptance of the spectrometer enables the study of a wide
range of reaction products but requires the reconstruction of numerous
correlated observables, making data analysis a highly multidimensional
problem.
}
\end{shaded*}
\label{fig:VAMOS}
\end{figure}

Recent advances in artificial intelligence (AI) and machine learning offer a new paradigm for 
experimental nuclear physics. By learning complex multidimensional correlations directly from 
data, AI-based methods can replace many of the most time-consuming analysis steps while 
maintaining or even improving precision. For VAMOS++, these developments have transformed 
data analysis from a major experimental bottleneck into a fast, reproducible, and scalable procedure.

Beyond accelerating existing analysis procedures, the VAMOS++ developments demonstrate how artificial 
intelligence can progressively transform experimental instrumentation itself, opening the path toward 
self-calibrating and ultimately self-optimizing detector systems.

This article reviews the challenges associated with VAMOS++ data analysis and describes how 
AI-based approaches are reshaping the way spectrometer data are analysed and interpreted.

\subsection{VAMOS++: A Powerful but Complex Instrument}

VAMOS++ (VAriable MOde Spectrometer) is a large-acceptance magnetic spectrometer designed 
to identify reaction products emitted in heavy-ion collisions. Its flexibility and broad acceptance 
make it suitable for a diverse range of experiments, from studies of exotic nuclei to investigations 
of nuclear dynamics and fission processes.

The spectrometer combines magnetic elements with a sophisticated focal-plane detection system. 
As ions travel through the magnetic field, their trajectories depend on several physical quantities, 
including magnetic rigidity, emission angles, position at the target, velocity, atomic number, mass number, 
and atomic charge state.  Figure~\ref{fig:VAMOS} shows  a schematic view of the VAMOS++ magnetic spectrometer and detection system.

\begin{shaded*}
\noindent {\bf What is a Deep Neural Network?}

\noindent A deep neural network (DNN) is a machine-learning model designed to learn complex relationships between input data and desired 
outputs. It consists of multiple layers of interconnected mathematical units ("neurons") that progressively transform information and 
identify patterns hidden within large datasets.

\noindent Unlike traditional analytical approaches, where the relationship between measured quantities and physical parameters must be 
explicitly formulated, a DNN learns these relationships directly from examples during a training phase. 

\noindent The main strength of DNNs is their ability to describe highly nonlinear and multidimensional systems. 

\noindent Importantly, DNNs do not replace physical understanding. Rather, they provide a powerful mathematical framework for extracting 
information from complex datasets.

\end{shaded*}

To determine these quantities, VAMOS++ records a large number of observables from multiple detectors. 
The final particle identification requires combining information from tracking detectors, timing measurements, 
energy-loss signals, residual energy detectors, and magnetic transport properties. Because these quantities 
are strongly correlated and often affected by detector imperfections and nonlinear effects, the analysis naturally 
becomes a multidimensional problem.

Historically, extracting the desired physical quantities has required successive analysis procedures, 
empirical corrections, and repeated iterations~\cite{Lemasson2023a}. 
Although highly successful, these approaches demand significant expertise and often represent a substantial fraction 
of the overall effort associated with an experimental campaign.

\subsection{Traditional Analysis Approaches and Their Limitations}

The conventional VAMOS++ analysis chain was developed over many years and has enabled numerous important 
scientific results. In this framework, each physical quantity is reconstructed through dedicated analytical procedures 
based on transport matrices, detector calibrations, and empirical corrections.

Trajectory reconstruction provides a representative example. The goal is to determine the ion properties at the 
target position from measurements performed at the entrance and the focal plane of VAMOS++. Because the magnetic 
transport through the spectrometer is highly nonlinear and depends on several coupled variables, the trajectory reconstruction 
requires sophisticated mathematical models and iterative optimization procedures.

\begin{shaded*}
\noindent {\bf Why are DNNs attractive for experimental physics?}

\begin{itemize}
\item They naturally handle many correlated variables simultaneously.
\item They can model nonlinear detector responses.
\item They require fewer simplifying assumptions than analytical parameterizations.
\item They process large datasets rapidly once trained.
\item They provide reproducible and automated analysis procedures.
\item They can be applied in supervised, weakly supervised, or unsupervised learning settings.
\end{itemize}
\end{shaded*}

Similarly, the determination of atomic charge state, atomic number, and mass number involves combining signals 
from several detector systems. Small imperfections in calibrations, detector responses, detector thicknesses, or trajectory 
reconstruction parameters can propagate through the analysis chain and affect the final particle identification.

Several challenges emerge from this traditional methodology:
\begin{itemize}
\item Analysis procedures can require months or even years of effort.
\item Numerous calibration and optimization steps must be repeated for each experiment.
\item Human intervention introduces unavoidable variability between analyses.
\item Increasing data volumes make manual approaches progressively less scalable.
\item The multidimensional nature of the problem limits the effectiveness of simplified analytical treatments.
\end{itemize}

These limitations do not arise from shortcomings of the spectrometer itself. Rather, they reflect the growing 
complexity of modern experiments and the difficulty of extracting information from increasingly rich datasets.

\subsection{Artificial Intelligence as a New Analysis Paradigm}

Machine learning methods provide a fundamentally different approach to the data analysis problem. 
Instead of relying on explicit analytical parameterizations of multidimensional detector responses, neural networks 
learn the relationship between measured detector signals and the desired physical quantities directly from experimental 
or simulated datasets.

\begin{figure}
\centering
\begin{shaded*}
\includegraphics[width=1\columnwidth]{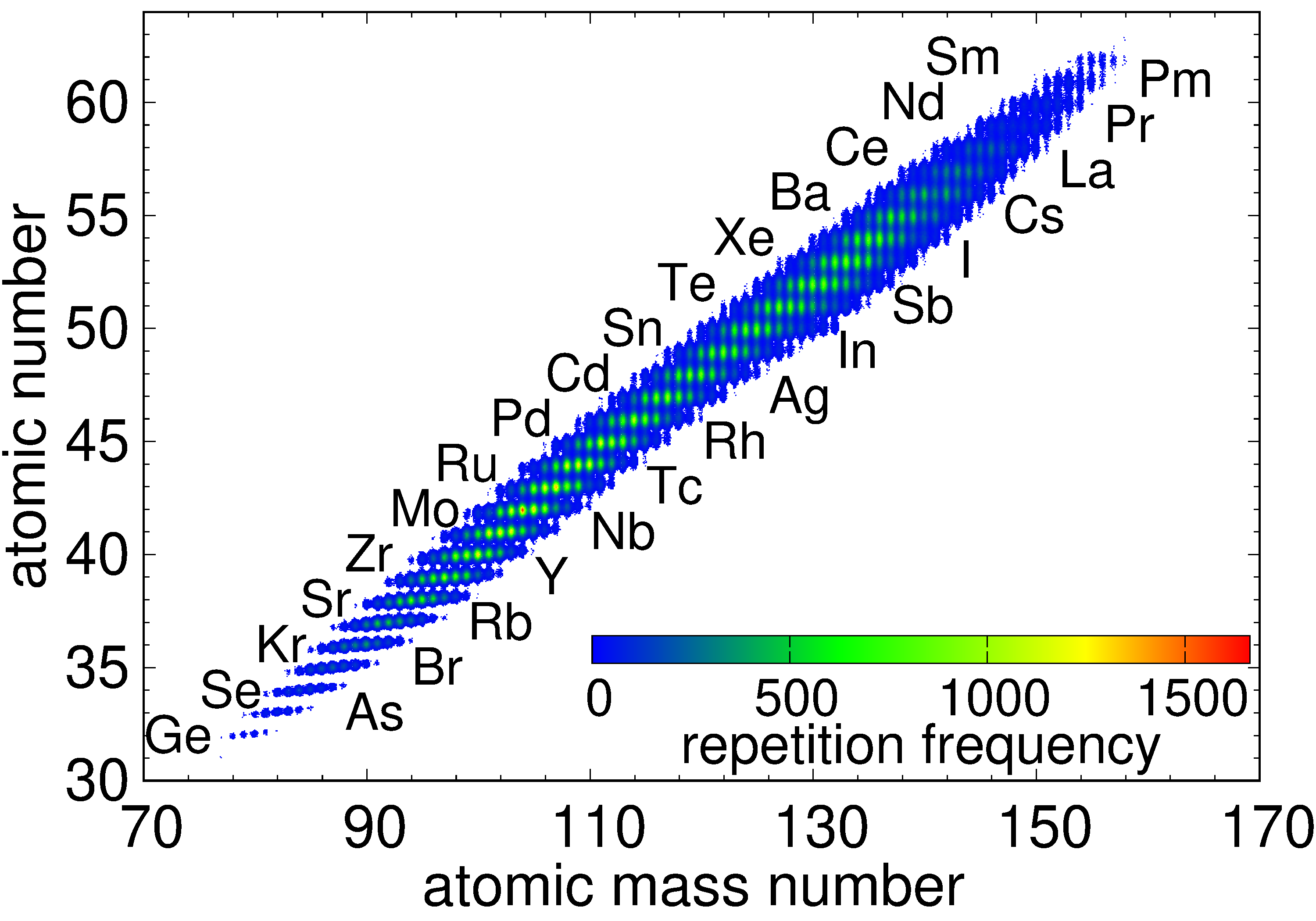}
\caption{
{\bf VAMOS++ isotopic identification:}
High-resolution event-by-event isotopic two-dimensional identification spectra for atomic number, and atomic mass 
number were obtained for each atom individually utilizing the VAMOS++ detector and artificial intelligence-based data analysis. 
These spectra correspond to the fission fragments detected in VAMOS++ and produced during the collisions of the $^{238}$U beam 
at an energy of $1399$~MeV with the $^{9}$Be target.
}
\end{shaded*}
\label{fig:RES}
\end{figure}

Over the past several years, a series of developments has demonstrated the effectiveness of AI-based methods for 
VAMOS++ analysis.

The first applications focused on the ion trajectory reconstruction~\cite{Rejmund2025a}. Deep neural-network models were trained 
to infer ion properties at the target from entrance and focal-plane detector measurements. By simultaneously 
treating multiple correlated variables, deep neural networks captured nonlinear dependencies that are difficult to describe 
using traditional approaches. The resulting models achieved excellent trajectory reconstruction accuracy while significantly 
accelerating the analysis process.

More recently, AI methods have been applied to the identification of atomic charge state and atomic number~\cite{Rejmund2025b}. 
A particularly important innovation was the use of fractionally labelled events, only a few lowest atomic charge states and atomic numbers
were labelled. This allows the learning algorithm to exploit information from partially ambiguous measurements. This approach 
improves classification performance, makes more efficient use of available experimental data and reduces the analysis time from 
months or years to several minutes.
\begin{shaded*}
\noindent {\bf Deep Neural Networks at VAMOS++}

\noindent For detector systems such as VAMOS++, the capability of DNN to discover multidimensional and nonlinear correlations 
directly from data is particularly valuable. The measured signals originate from multiple detectors and are influenced 
simultaneously by many correlated variables, and instrumental effects. 

At VAMOS++, DNNs are currently applied to three classes of problems:
\begin{itemize}
\item trajectory reconstruction through the magnetic spectrometer,
\item particle identification (atomic charge state, atomic number, and mass number),
\item detector calibration and response correction.
\end{itemize}
\end{shaded*}

Beyond their excellent performance, AI-based methods provide several practical advantages:

\begin{itemize}
\item Rapid processing of large experimental datasets.
\item Consistent and reproducible analysis procedures.
\item Reduced dependence on manual tuning and expert intervention.
\item Improved treatment of multidimensional correlations.
\item Robust handling of nonlinear detector responses.
\item Easy adaptation to future detector upgrades and new experimental conditions.
\end{itemize}

Ultimately, the principal benefit of AI is not faster computation but the ability for physicists to devote 
more effort to understanding nuclear phenomena rather than performing repetitive analysis 
and calibration procedures.

\subsection{Results and Scientific Impact}

The AI-based analysis methods developed for VAMOS++ have demonstrated performance comparable to 
or better than traditional approaches across multiple analysis tasks.

Trajectory reconstruction achieves unprecedented high precision while processing millions of event per second.

Atomic charge-state and atomic-number identification benefit from improved classification capabilities and
enhanced robustness in regions where conventional methods become challenging while reducing processing times 
by orders of magnitude.  Figure~\ref{fig:RES} illustrates the VAMOS++ isotopic identification performances reached using AI-based analysis methods.

These advances directly benefit the scientific programs supported by VAMOS++. Faster and more reliable analysis 
allows experimental collaborations to process larger datasets, explore broader regions of parameter space, and 
obtain physics results more rapidly. The improved reproducibility also facilitates long-term comparisons between 
experiments and promotes greater consistency across analyses performed by different users.

The impact extends beyond VAMOS++ itself. The methodologies developed for trajectory reconstruction and particle 
identification are broadly applicable to other spectrometers and detector systems facing similar multidimensional 
reconstruction challenges. As experimental facilities continue to generate increasingly large and complex datasets, 
AI-based analysis techniques are expected to become an essential component of the nuclear-physics toolkit.

\begin{figure}
\centering
\begin{shaded*}
\includegraphics[width=1\columnwidth]{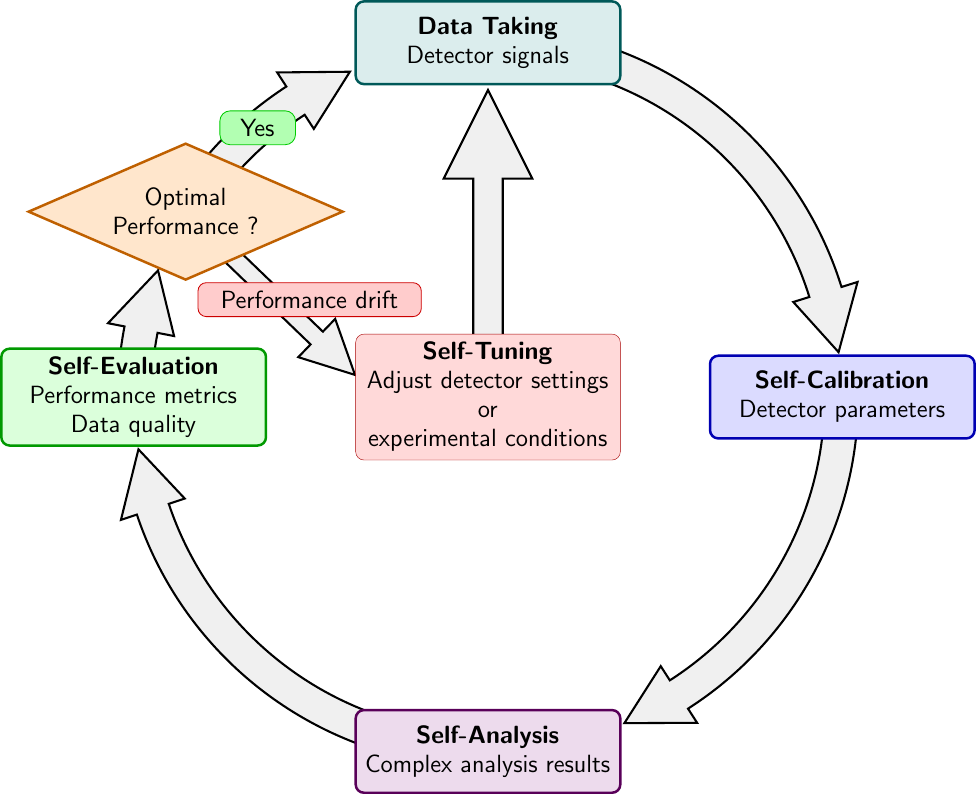}
\caption{
{\bf Future intelligent instrumentation:}
Conceptual vision of an AI-driven adaptive experimental workflow for
intelligent instrumentation. Data taking, detector calibration, data analysis, 
and performance monitoring are integrated into a continuous feedback loop. 
The system evaluates its own performance and, when necessary, automatically adjusts detector settings or 
experimental conditions before continuing data acquisition. 
}
\end{shaded*}
\label{fig:FUTURE}
\end{figure}

\subsection{Perspectives: Towards Self-Calibrating Nuclear Instrumentation}

The developments achieved so far for VAMOS++ represent only the first stage of a broader transformation in experimental 
nuclear physics. Initial applications focused on improving and accelerating trajectory reconstruction and particle identification, 
demonstrating that machine-learning methods can reproduce and often improve upon traditional analysis procedures while 
dramatically reducing the required effort and processing time.

A particularly important challenge in spectrometer analysis is detector calibration. Conventional calibration procedures 
typically require dedicated datasets, extensive expert intervention, and multiple iterations before a satisfactory solution 
is obtained. In many cases, subsequent analysis steps depend critically on the quality of these calibrations.

In previous work on atomic charge-state and atomic-number identification, machine-learning algorithms were trained 
using fractionally labelled events. While highly successful, this approach still relied on an approximate online calibration 
of the ionization chamber signals as a starting point for the analysis.

Current developments aim to remove even this requirement. New unsupervised deep-learning methods are being developed 
to determine the calibration coefficients of the segmented ionization chamber directly from experimental data. By exploiting 
the raw energy signals together with independently reconstructed mass-over-charge information derived from time-of-flight 
measurements and trajectory reconstruction, the network can infer the detector calibration parameters without the need for 
pre-labelled calibration datasets. The procedure therefore becomes largely autonomous and fully data-driven.

Beyond the ionization chamber, similar concepts are being explored for the Multi-Wire Parallel Plate Avalanche Counters 
(MWPPACs), which provide the tracking information required for trajectory reconstruction. Ongoing work focuses on 
deep-learning approaches capable of simultaneously calibrating detector responses, determining charge-sharing characteristics, 
and reconstructing particle positions while enforcing physically motivated constraints. Such methods have the potential to 
reduce calibration effort substantially while improving consistency and reproducibility across experiments.

These developments point toward a future in which detector calibration, signal reconstruction, particle identification, and trajectory 
reconstruction are no longer treated as separate tasks. Instead, they may become components of a unified analysis framework that 
continuously learns from experimental data and automatically adapts to changing detector conditions.

The evolution of AI applications at VAMOS++ can be viewed as a sequence of increasing autonomy: first accelerating trajectory 
reconstruction, then automating particle identification, next enabling self-calibration, and ultimately leading toward self-optimizing 
instrumentation. Each step reduces the need for manual intervention and moves experimental systems closer to continuous 
autonomous operation.

The long-term implications extend beyond self-calibration. Once detector calibration and reconstruction become data-driven and largely 
autonomous, the possibility emerges for self-optimizing instrumentation. Rather than periodically correcting detector responses through 
dedicated calibration procedures, future systems could continuously monitor detector performance and automatically adjust reconstruction 
and calibration parameters as experimental conditions evolve. Figure~\ref{fig:FUTURE} illustrates such an AI-driven adaptive experimental workflow for intelligent instrumentation.

For spectrometers such as VAMOS++, such adaptive approaches could compensate for detector aging, gain drifts, changing beam 
conditions, or variations in experimental environments in near real time. The analysis framework would no longer be a passive tool 
applied after data acquisition, but an active component of the experimental system itself.

Looking further ahead, one can envision intelligent detector systems in which calibration, reconstruction, performance monitoring, 
and quality assurance operate within a unified machine-learning framework. Such systems could continuously evaluate their own 
performance, identify anomalies, suggest corrective actions, and maintain optimal operating conditions throughout an experimental campaign.

While these developments remain a subject of ongoing research, the recent progress achieved with AI-based reconstruction and 
unsupervised calibration at VAMOS++ suggests that this vision is becoming increasingly realistic. In this sense, artificial intelligence 
may ultimately transform not only how experimental data are analyzed, but also how complex nuclear-physics instruments are operated.

\subsection{Conclusions}
Artificial intelligence has rapidly evolved from a promising computational tool into a practical component of the VAMOS++ analysis 
framework. Applications ranging from trajectory reconstruction to particle identification have demonstrated substantial gains in 
precision, reproducibility, and efficiency. Current developments in unsupervised detector calibration further extend the role of AI 
from data analysis toward instrumentation itself. Looking ahead, the combination of self-calibration, continuous performance 
monitoring, and adaptive optimization may enable a new generation of intelligent nuclear-physics instruments capable of maximizing 
scientific output while minimizing analysis overhead. The VAMOS++ program illustrates how AI can help experimentalists spend 
less time analyzing data and more time understanding the underlying physics.
 
\bibliographystyle{apsrev4-2}
\bibliography{NPN.bib}   

\end{document}